\documentclass[11pt]{article}

\usepackage[a4paper,margin=2.5cm]{geometry}
\usepackage{amsmath,amssymb}
\usepackage{graphicx}
\usepackage{bm}
\usepackage{hyperref}
\usepackage{authblk}
\usepackage{newtxtext}
\usepackage{newtxmath}
\usepackage[all]{hypcap}
\usepackage{multirow}
\usepackage{booktabs}
\usepackage{bbold}
\usepackage{bm}
\usepackage{bbm}
\newcommand{\ket}[1]{|{#1}\rangle}

\hypersetup{
	colorlinks=true,
	linkcolor=blue,
	citecolor=blue,
	urlcolor=blue
}

\title{\bf High-resolution wide-field magnetic imaging with sparse sampling using nitrogen-vacancy centers}

\author[1,2,3]{Keqing Liu}
\author[1,2,3,*]{Jiazhao Tian}
\author[1,2,3]{Bokun Duan}
\author[1,2,3]{Hao Zhang}
\author[1,2,3]{Kangze Li}
\author[4]{Guofeng Zhang}
\author[5]{Fedor Jelezko}
\author[5]{Ressa S. Said}
\author[6,7]{Jianming Cai}
\author[1,2,3,4,*]{Liantuan Xiao}

\affil[1]{College of Physics and Optoelectronic Engineering, Taiyuan University of Technology, Taiyuan 430000, People's Republic of China}
\affil[2]{Shanxi Key Laboratory of Precision Measurement Physics, Taiyuan University of Technology, Taiyuan 030024, China}
\affil[3]{Key Laboratory of Advanced Transducers and Intelligent Control System, Ministry of Education, and Shanxi Province, Taiyuan University of Technology, Taiyuan 030024, China}
\affil[4]{State Key Laboratory of Quantum Optics Technologies and Devices, Institute of Laser Spectroscopy, Collaborative Innovation Center of Extreme Optics, Shanxi University, Taiyuan, 030006, China}
\affil[5]{Institute for Quantum Optics \& Center for Integrated Quantum Science and 
	Technology, Universit\"{a}t Ulm, 89081 Ulm, Germany}
\affil[6]{School of Physics, Hubei Key Laboratory of Gravitation and Quantum Physics, Institute for Quantum Science and Engineering, International Joint Laboratory on Quantum Sensing and Quantum Metrology, Center for Intelligence and Quantum Science (CIQS), Huazhong University of Science and Technology, Wuhan 430074, China}
\affil[7]{Wuhan Institute of Quantum Technology, Wuhan 430074, China}

\begin{document}
	
	\maketitle
	
	\begin{center}
		\small
		\texttt{*tianjiazhao@tyut.edu.cn}, 
		\texttt{*xlt@sxu.edu.cn}
	\end{center}
	
	\begin{abstract}
		Nitrogen--vacancy (NV) centers in diamond enable quantitative magnetic imaging, yet practical implementations must balance spatial resolution against acquisition time (and thus per-pixel sensitivity). Single-NV scanning magnetometry achieves genuine nanoscale resolution, nonetheless requires typically a slow pixel-by-pixel acquisition. Meanwhile, wide-field NV-ensemble microscopy provides parallel readout over a large field of view, however is jointly limited by the optical diffraction limit and the sensor--sample standoff. Here, we present a sparse-sampling strategy for reconstructing high-resolution wide-field images from only a small number of measurements. Using simulated NV-ensemble detection of ac~magnetic fields, we show that a mean-adjusted Bayesian estimation (MABE) framework can reconstruct $10^{4}$-pixel images from only 25 sampling points, achieving SSIM values exceeding 0.999 for representative smooth field distributions, while optimized dynamical-decoupling pulse sequences yield an approximately twofold improvement in magnetic-field sensitivity. The method further clarifies how sampling patterns and sampling density affect reconstruction accuracy and suggests a route toward faster and more scalable magnetic-imaging architectures that may extend to point-scanning NV sensors and other magnetometry platforms, such as SQUIDs, Hall probes, and magnetic tunnel junctions.
		
	\end{abstract}

	\section{Introduction}
	
	Nitrogen--vacancy (NV) centers in diamond provide a well-defined electronic spin system in a chemically robust solid-state host and have therefore become a central platform for quantum sensing and metrology~\cite{Degen2017Quantumsensing,Doherty2013nitrogenvacancycolourcentre,Schirhagl2014NitrogenVacancyCentersDiamond}. In these centers, the ground-state spin sublevels shift in response to external perturbations and can be read out optically, enabling quantitative measurements of magnetic fields via shifts of the spin resonance frequencies detected by optically detected magnetic resonance (ODMR)~\cite{Bucher2019Quantumdiamondspectrometer}, electric fields through Stark shifts~\cite{Dolde2011Electricfieldsensingusing}, microwave-frequency magnetic fields via driven spin transitions~\cite{Aslam2017Nanoscalenuclearmagnetic}, as well as temperature through the temperature dependence of the zero-field splitting parameter~\cite{Acosta2010TemperatureDependenceNitrogenVacancy,Kucsko2013Nanometrescalethermometryliving,Doherty2013nitrogenvacancycolourcentre}, and lattice strain via strain-induced shifts and splittings of the NV spin levels~\cite{Doherty2013nitrogenvacancycolourcentre,Broadway2019MicroscopicImagingStress,Kehayias2019Imagingcrystalstress}, by means of optically detected magnetic resonance techniques. Within this general framework, NV-based magnetometry has emerged as one of the most mature sensing modalities, with applications ranging from nanoscale studies of condensed-matter systems~\cite{Casola2018Probingcondensedmatter} and imaging of geological and mineral samples using quantum-diamond microscopy~\cite{Levine2019Principlestechniquesquantum}, to biological signal detection and biomedical contexts~\cite{WOS:000628193200001}, as well as diagnosis and fingerprinting of activity in integrated circuits~\cite{Turner2020MagneticFieldFingerprinting,Webb2022HighSpeedWideFieldImaginga}.
	
	For spatially resolved measurements, magnetic-field imaging places simultaneous demands on spatial resolution and on the acquisition time required to reach a target field sensitivity per pixel, reflecting the standard $1/\sqrt{T}$ scaling of sensitivity with averaging time in NV magnetometry~\cite{Barry2020SensitivityoptimizationNVdiamond,Scholten2021Widefieldquantummicroscopy}. Scanning-probe approaches based on single, shallow NV centers attached to atomic-force or planar probes can achieve genuine nanoscale spatial resolution by maintaining sensor--sample separations of only a few tens of nanometres, thereby accessing the near-field stray-field distribution of the sample~\cite{Hong2013NanoscalemagnetometryNV,Rovny2022Nanoscalecovariancemagnetometry,WOS:001366175800001}. In practice, however, each pixel typically requires dwell times on the order of milliseconds or longer to obtain sufficient signal-to-noise ratio, so that high-resolution images are restricted to fields of view of at most a few to a few tens of square micrometres and acquisition times of minutes or more~\cite{Hong2013NanoscalemagnetometryNV,Rovny2022Nanoscalecovariancemagnetometry,Levine2019Principlestechniquesquantum}.
	
	An alternative architecture employs near-surface NV ensembles interrogated in a widefield optical microscope, where NV fluorescence is detected on a camera to form spatially resolved maps of local magnetic fields in parallel across the full diamond area~\cite{Levine2019Principlestechniquesquantum,Scholten2021Widefieldquantummicroscopy}. Such quantum diamond microscopes typically offer fields of view ranging from tens of micrometres up to the millimetre scale while retaining quantitative magnetometric performance and mitigating artefacts associated with mechanical drift of scanning probes~\cite{Levine2019Principlestechniquesquantum,Webb2022HighSpeedWideFieldImaginga}. Because the four crystallographic NV orientations can be addressed separately, ensemble-based widefield platforms also enable vector reconstruction of the local magnetic field and support measurements at or near zero bias field~\cite{Scholten2021Widefieldquantummicroscopy,WOS:000665592700001,WOS:000930619100001,WOS:000665534100001}.
	
	A key limitation of conventional widefield NV microscopy is that the lateral spatial resolution is jointly determined by the sensor--sample standoff and the diffraction limit of far-field imaging optics, such that even with high-numerical-aperture objectives the effective resolution typically remains in the few-hundred-nanometre range~\cite{Levine2019Principlestechniquesquantum,Scholten2021Widefieldquantummicroscopy}. To relax this constraint, several super-resolution strategies have been adapted to NV centers, including stimulated-emission-depletion microscopy of single NVs in bulk diamond and nanodiamonds~\cite{Han2009ThreeDimensionalStimulatedEmission,Arroyo-Camejo2013StimulatedEmissionDepletion}, charge-state-depletion schemes based on nonlinear control of NV$^{-}$/NV$^{0}$ charge conversion~\cite{Chen2015Subdiffractionopticalmanipulation}, and more recently super-resolution-enabled widefield quantum-diamond microscopy using structured illumination~\cite{Xu2024SuperResolutionEnabledWidefield}. These techniques can push the effective spatial resolution into the $\sim10$--$100~\mathrm{nm}$ regime, but most implementations still rely on scanning or repeated patterned excitation and therefore do not yet offer the acquisition speed and simplicity of conventional widefield NV microscopy~\cite{Scholten2021Widefieldquantummicroscopy,Chen2015Subdiffractionopticalmanipulation,Xu2024SuperResolutionEnabledWidefield}.
	
	In this work, we propose a sparse-sampling strategy that enables high-resolution widefield imaging from a limited number of measurement points. Using simulated NV ensemble measurements of ac magnetic fields as an illustrative example, we demonstrate that by sampling only $25$ points within the field of view and applying a mean-adjusted Bayesian estimation (MABE) scheme~\cite{Stuerner2021PortableNV,xie_microfabricated_2022,homrighausen_microscale_2024,basso_wide-field_2025,ku_imaging_2020,tang_quantum_2023}, a high-resolution image containing $10^{4}$ pixels can be reconstructed. For magnetic-field patterns comprising different numbers of extrema, the reconstructed images achieve structural similarity index (SSIM) values exceeding $0.999$. Moreover, by optimizing the dynamical-decoupling pulse sequence, the magnetometry contrast is enhanced, leading to an approximate twofold improvement in magnetic-field sensitivity. We further analyze how sampling strategies and sampling density affect reconstruction quality, providing guidance for adapting the method to different application requirements.
	
	As a general sparse-sampling imaging framework, the proposed MABE approach can be straightforwardly extended to point-scanning magnetometry platforms, including single-NV-based schemes as well as other sensor technologies such as superconducting quantum interference devices (SQUIDs), Hall sensors, and magnetic tunnel junctions. These results point toward fast, compact, and scalable quantum magnetic imaging architectures with potential applications in on-chip magnetic diagnostics, real-time field mapping, and integrated quantum metrology~\cite{Stuerner2021PortableNV,xie_microfabricated_2022,homrighausen_microscale_2024,basso_wide-field_2025,ku_imaging_2020,tang_quantum_2023}.
	
	\section{AC Magnetic Field Measurement}
	
	The negatively charged nitrogen-vacancy (NV$^{-}$) center in diamond possesses an electron spin $S=1$ ground state with a zero-field splitting $D = 2\pi\times2.88~\mathrm{GHz}$~\cite{jelezkoSingleDefectCentres2010}. When driven by a sequence of microwave pulses that flip the spin state at each zero crossing of an ac magnetic field, the field amplitude can be extracted from the accumulated spin phase. We consider a two-level subspace spanned by $\ket{0}$ and $\ket{1}$ (representing $\ket{+1}$ or $\ket{-1}$) and denote the actual frequency of microwave as $\omega$. In the interaction picture with respect to $\omega\sigma_z$, the Hamiltonian is
	\begin{equation}
		H = \frac{\delta_0}{2}\sigma_z + H_{\mathrm{AC}} + H_{\mathrm{C}},
	\end{equation}
	where $\delta_0$ is the detuning between the microwave frequency and the $\ket{0}\leftrightarrow\ket{1}$ transition, $H_{\mathrm{AC}} = g_{\mathrm{AC}}\cos(\omega_{\mathrm{AC}} t)\sigma_z$ describes the ac magnetic field, and $H_{\mathrm{C}} = \Omega_x(t)\sigma_x + \Omega_y(t)\sigma_y$ represents the microwave control pulses.
	
	In the ideal noiseless case, $\pi$ pulses are applied at each zero crossing of the ac field, corresponding to a modulation function $f(t)=\mathrm{sgn}[\sin(\omega_{\mathrm{AC}} t)]$. The accumulated relative phase is~\cite{Degen2017Quantumsensing,genovEfficientRobustSignal2020,Ishikawa2018DDwidth}
	\begin{equation}
		\Phi(t) = 2\int_0^t g_{\mathrm{AC}}\cos(\omega_{\mathrm{AC}} t')f(t')\,dt' \approx \frac{4}{\pi}g_{\mathrm{AC}}t.
	\end{equation}
	After a Ramsey measurement, the population of $\ket{0}$ follows $P_0=\frac{1}{2}[1+\cos\Phi(t)]$. The magnetic-field amplitude is given by $B=g_{\mathrm{AC}}/\gamma=\pi\omega/(4\gamma)$, where $\gamma$ is the gyromagnetic ratio of the NV center.
	
	\begin{figure}
		\centering
		\includegraphics[width=0.9\textwidth]{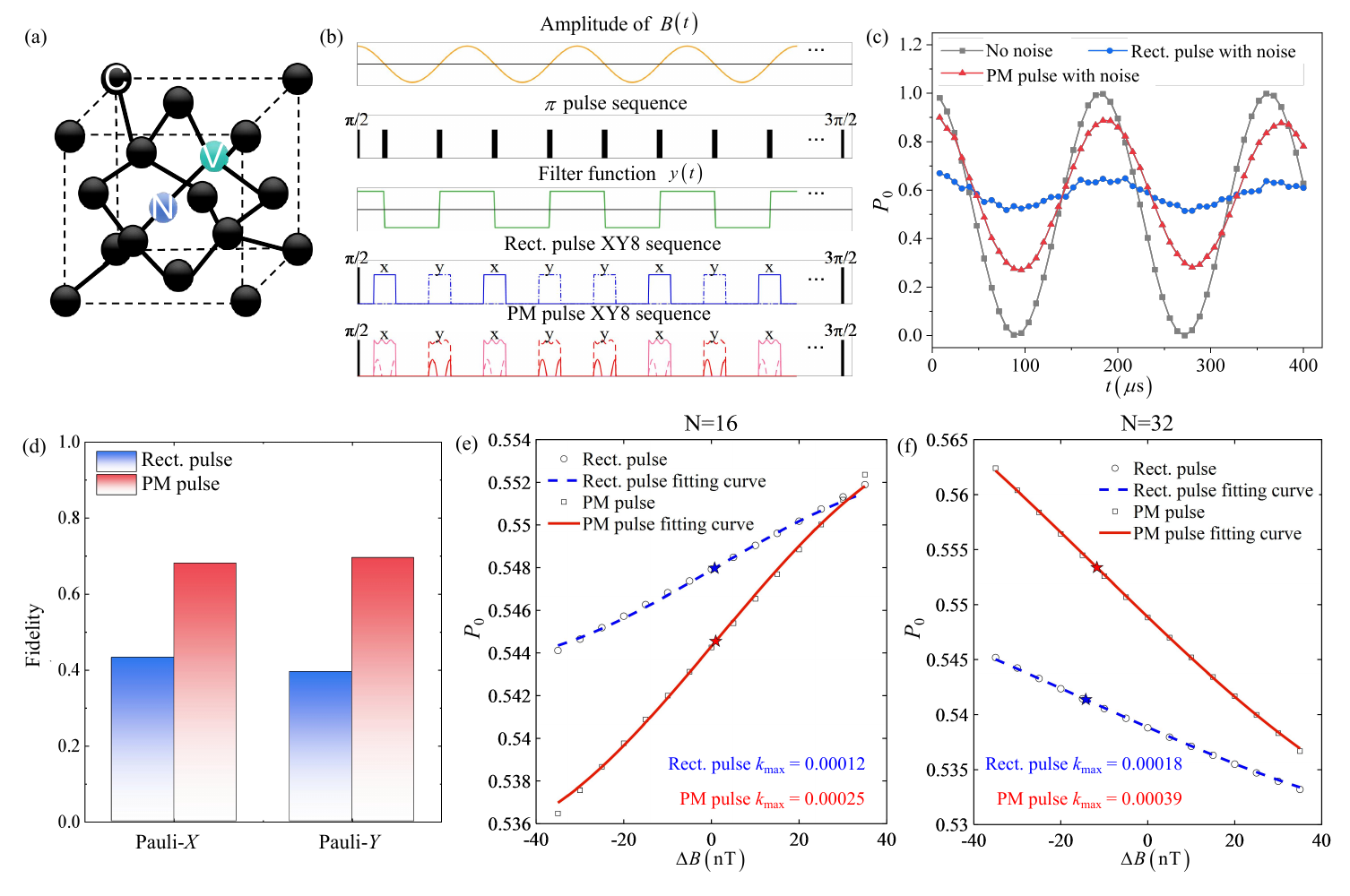}
		\caption{(a) Schematic illustration of an NV center in the diamond lattice.
			(b) Schematic representation of the temporal correspondence among the target AC magnetic field, the control pulses, and the filter function. From top to bottom: amplitude of the target AC magnetic field; ideal infinite-duration pulse sequence applied at the zero crossings; the corresponding filter function for the ideal pulse sequence; XY-8 rectangular-pulse sequence with finite pulse width; and XY-8-PM sequence with finite pulse width. In our simulation, we set the pulse duration $T_{\mathrm{pulse}}=50$~ns, the pulse interval $\tau_{p}=950$~ns, and the total evolution time $t = 8N\left(T_{\mathrm{pulse}}+\tau_{p}\right)=400~\mu$s, with the sequence number $N = 50$.
			(c) Time evolution of the NV spin population. The gray line with square markers corresponds to the noiseless case using ideal infinite pulses, whereas the blue line with dot markers and the red line with triangular markers correspond to rectangular pulses and PM pulses, respectively, in the presence of inhomogeneous broadening noise and the dynamical noise (see the main text).
			(d) Comparison of the Pauli-$X$ and Pauli-$Y$ gate fidelities for rectangular pulses and PM pulses, with fidelity defined in Eq.~(7) and PM pulse forms given in Eqs.~(8)--(9).
			(e) Dependence of the normalized spin population for $N=16$ ($t=128~\mu$s). The starred markers denote the points of maximum slope: $k=0.00012$ for rectangular pulses and $k=0.00025$ for PM pulses.
			(d) Dependence of the normalized spin population for $N=32$ ($t=256~\mu$s). The starred markers denote the points of maximum slope: $k=0.00018$ for rectangular pulses and $k=0.00039$ for PM pulses.}
		\label{Fig1}
	\end{figure}
	
	In practical NV ensemble measurements, inhomogeneous broadening induced by nonuniform local environments reduces the fidelity of $\pi$ pulses, thereby degrading the spin-population contrast and ultimately the magnetic-field sensitivity. We decompose the detuning term as $\delta_0 = \delta + \delta_d(t)$, where $\delta$ represents time-independent inhomogeneous broadening described by a Gaussian distribution~\cite{PhysRevLett.115.190801}
	\begin{equation}
		p(\delta)=\frac{1}{\sqrt{2\pi}\sigma_g}\exp\left(-\frac{\delta^2}{2\sigma_g^2}\right),
	\end{equation}
	and $\delta_d(t)$ denotes time-dependent noise following an Ornstein--Uhlenbeck process~\cite{RevModPhys.17.323,PhysRevE.54.2084}
	\begin{equation}
		\delta_d(t+\Delta t)=\delta_d(t)e^{-\Delta t/\tau}
		+\left[\frac{c\tau}{2}\left(1-e^{-2\Delta t/\tau}\right)\right]^{1/2}n_{\delta},
	\end{equation}
	where $\tau$ and $c$ are the correlation time and diffusion constant, respectively, and $n_{\delta}\sim\mathcal{N}(0,1)$. We set the full width at half maximum (FWHM) of the static detuning distribution to $W=2\sqrt{2\ln2}\,\sigma_g=2\pi\times26.5~\mathrm{MHz}$, the correlation time to $\tau=20~\mu\mathrm{s}$, and the diffusion coefficient to $\sqrt{c\tau/2}=2\pi\times50~\mathrm{kHz}$~\cite{genovEfficientRobustSignal2020}. The pulse duration is fixed at $T_{\mathrm{pulse}}=50~\mathrm{ns}$, corresponding to a Rabi frequency of $10~\mathrm{MHz}$, with an interpulse interval $\tau_p=950~\mathrm{ns}$ and a total evolution time $t=8N(T_{\mathrm{pulse}}+\tau_p)=400~\mu\mathrm{s}$ where the sequence number $N=50$.
	
	Under these noise conditions, the spin-population evolution obtained using an XY-8 sequence with rectangular $\pi$ pulses is shown in Fig.~\ref{Fig1}(c) (blue curve with dot markers), where a pronounced reduction in contrast to below $0.2$ is observed.
	
	To mitigate this effect, we employ a phase-modulated (PM) pulse optimization strategy~\cite{tianQuantumOptimalControl2020} to enhance pulse fidelity and thereby improve population contrast and magnetic-field sensitivity. The PM control Hamiltonian is expressed as
	\begin{equation}
		H_{\mathrm{PM}}(t)=\sum_{j=1}^{N}\frac{a_j}{2}
		\left\{
		\cos\!\left[\frac{b_j}{\nu_j}\sin(\nu_j t)\right]\sigma_x
		+\sin\!\left[\frac{b_j}{\nu_j}\sin(\nu_j t)\right]\sigma_y
		\right\},
	\end{equation}
	which generates a multifrequency waveform with a reduced parameter space. The parameters $\{a_j,b_j,\nu_j\}$ are optimized by maximizing the objective function
	\begin{equation}
		F_{\mathrm{obj}}^G=\mathcal{N}\sum_{k=1}^{M}p(\delta_k)f_g(\delta_k),
	\end{equation}
	where $\mathcal{N}=[\sum_k p(\delta_k)]^{-1}$ is a normalization factor and
	\begin{equation}
		f_g(\delta)=\frac{1}{2}+\frac{1}{3}\sum_{\kappa=x,y,z}
		\mathrm{Tr}\!\left(
		U_{\mathrm{tar}}\frac{\sigma_\kappa}{2}U_{\mathrm{tar}}^\dagger
		U_\delta\frac{\sigma_\kappa}{2}U_\delta^\dagger
		\right)
	\end{equation}
	is the average gate fidelity~\cite{BOWDREY2002258,NIELSEN2002249}. Here
	$U_\delta=T\exp[-i\int_0^t H(\delta,t')dt']$ and the target operators
	$U_{\mathrm{tar}}$ correspond to Pauli-$X$ and Pauli-$Y$ gates for
	$\pi_x$ and $\pi_y$ pulses, respectively.
	
	We uniformly sample $M=15$ detuning values within the interval $[-W,W]$, covering approximately $98\%$ of the distribution. The pulse amplitude is constrained below $2\pi\times10~\mathrm{MHz}$. The optimization can be carried out using standard implementations of direct-search optimization methods.
	
	The resulting improvements in Pauli-$X$ and Pauli-$Y$ gate fidelities are shown in Fig.~\ref{Fig1}(d), where both increase from approximately $0.43$ to $0.68$. To preserve orthogonality between $\pi_x$ and $\pi_y$ pulses in the XY-8 sequence, identical PM parameters are used. We adopt the parameter set $a=0.0628$, $b=0.0830$, and $\nu=0.0316$, yielding the control Hamiltonians
	\begin{equation}
		H_{\pi_x}(t)=\frac{a}{2}\!\left[
		\cos\!\left(\frac{b}{\nu}\sin\nu t\right)\sigma_x
		+\sin\!\left(\frac{b}{\nu}\sin\nu t\right)\sigma_y
		\right],
	\end{equation}
	and
	\begin{equation}
		H_{\pi_y}(t)=\frac{a}{2}\!\left[
		\cos\!\left(\frac{b}{\nu}\sin\nu t\right)\sigma_y
		-\sin\!\left(\frac{b}{\nu}\sin\nu t\right)\sigma_x
		\right].
	\end{equation}
	
	As shown in Fig.~\ref{Fig1}(c), the optimized XY-8 sequence employing PM pulses enhances the population contrast to approximately $0.6$ in the presence of noise, representing a threefold improvement over the unoptimized case. This contrast enhancement directly translates into improved magnetic-field sensitivity~\cite{Pham2012EnhancedDD,genovEfficientRobustSignal2020}
	\begin{equation}
		\eta=\frac{\sigma}{(\partial P/\partial B)}\sqrt{T},
	\end{equation}
	where $\sigma$ is the standard deviation of single-shot readout noise which typically arises from the photon shot noise, $\partial P/\partial B$ is the maximum slope of the population response to magnetic-field amplitude, and $T$ is the measurement duration. Evaluating $\partial P/\partial B$ at $N=16$ and $N=32$, corresponding to evolution times $t=128~\mu\mathrm{s}$ and $256~\mu\mathrm{s}$, yields more than a twofold increase in slope (Fig.~\ref{Fig1}(e,f)), leading to sensitivity improvements from $0.94$ nT/$\sqrt{\text{Hz}}$ to $0.45$ nT/$\sqrt{\text{Hz}}$ and from $1.33$ nT/$\sqrt{\text{Hz}}$ to $0.64$ nT/$\sqrt{\text{Hz}}$, respectively, where $\sigma$ was taken to be $10^{-2}$\cite{Pham2012EnhancedDD}.
	
	\section{Imaging with the MABE Method}
	\begin{figure}[htbp]
		\centering
		\includegraphics[width=0.9\textwidth]{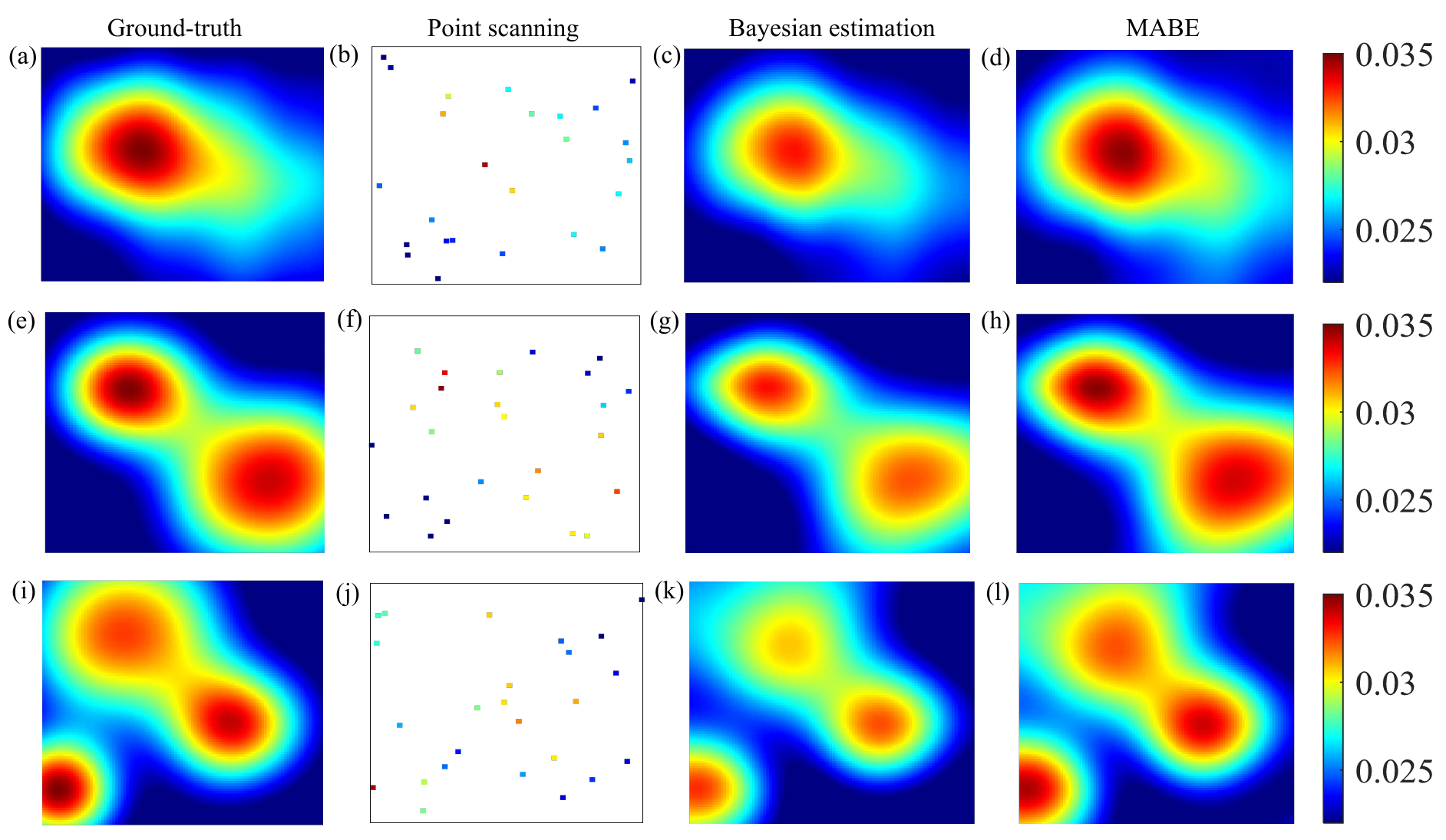}
		\caption{Imaging performance for magnetic-field distributions with different numbers of extrema.
			(a),(e),(i) Simulated ground-truth magnetic-field maps containing single, two, and three extrema.
			(b),(f),(j) Measurement results at $25$ randomly distributed sampling points (marker size enlarged for clarity).
			(c),(g),(k) Magnetic-field maps reconstructed directly from the sampled data using Bayesian estimation method, yielding $10^{4}$-pixel images.
			(d),(h),(l) Magnetic-field maps reconstructed from the same sampled data using the MABE method, also producing $10^{4}$-pixel images.}
		\label{Fig2}
	\end{figure}
	
	In this section, we introduce the mean-adjusted Bayesian estimation (MABE) method that enables rapid imaging of the magnetic field distribution. High-resolution magnetic field imaging typically relies on dense point-by-point scanning, which is highly time-consuming. In contrast, Bayesian-based estimation methods~\cite{Tian2023BayesHybridNV} allow the full distribution to be inferred from only a small number of samples. We denote $\mathbf{x}$ as the two-dimensional coordinate, $y(\mathbf{x})$ as the measured magnetic field value at $\mathbf{x}$, $\hat{y}(\mathbf{x})$ as the predicted magnetic field value, and $\mathbf{s}$ as the coordinates of the actual measurement samples. 
	The correlation between the errors at two coordinates $\mathbf{x}_i$ and $\mathbf{x}_j$ is defined as
	$Corr\left[\epsilon(\mathbf{x}_i), \epsilon(\mathbf{x}_j)\right] = \exp\left[-d\left(\mathbf{x}_i,\mathbf{x}_j\right)\right]$~\cite{1989Design},
	where $d$ is the distance function whose specific form is to be determined. The best linear unbiased predictor for $y(\mathbf{x})$ can be expressed as~\cite{jonesEfficientGlobalOptimization1998}
	
	\begin{equation}
		\hat{y}\left(\mathbf{x}\right)=\hat{\mu}+\mathbf{r}^{\prime}\left(\mathbf{x}\right)\mathbf{R}^{-1}\left(\mathbf{y}\left(\mathbf{s}\right)-\hat{I}\hat{\mu}\right),
	\end{equation}
	where $\mathbf{R}$ is an $n \times n$ correlation matrix with elements $Corr\left[\epsilon(\mathbf{s}_i), \epsilon(\mathbf{s}_j)\right]$, 
	$r\left(\mathbf{x}\right)=\left[R\left(\mathbf{s}_1,\mathbf{x}\right), \ldots, R\left(\mathbf{s}_n,\mathbf{x}\right)\right]^{\prime}$ is an $n \times 1$ vector representing the correlation between the errors of the sampling points and the untested input $\mathbf{x}$, with $R\left(\mathbf{s}_i,\mathbf{x}\right)=Corr[\epsilon(\mathbf{s}_i), \epsilon(\mathbf{x})]$, 
	$\hat{\mu}=\left(\hat{I}^{\prime}\mathbf{R}^{-1}\hat{I}\right)^{-1}\hat{I}^{\prime}\mathbf{R}^{-1}\mathbf{y}(\mathbf{s})$, 
	and $\hat{I}$ is the $n \times 1$ vector with all elements equal to $1$. 
	We take the distance function in the form
	$d\left(\mathbf{s}_i,\mathbf{s}_j\right)=\sum_{h=1}^{2}\alpha_h\left|s_{i,h}-s_{j,h}\right|^{P_h}$,
	with $\alpha_h \geq 0$ and $P_h \in [1,2]$, where the parameters $\alpha_h$ and $P_h$ are optimized by maximizing the likelihood function of the samples described by~\cite{jonesEfficientGlobalOptimization1998}
	
	\begin{equation}
		\frac{1}{\left(2\pi\right)^{n/2}\left(\sigma^{2}\right)^{n/2}\left|\mathbf{R}\right|^{1/2}}
		\exp\left[
		-\frac{\left(\mathbf{y}(\mathbf{s})-\hat{I}\mu\right)^{\prime}\mathbf{R}^{-1}
			\left(\mathbf{y}(\mathbf{s})-\hat{I}\mu\right)}{2\sigma^{2}}
		\right],
	\end{equation}
	where $\left|\mathbf{R}\right|$ denotes the determinant of $\mathbf{R}$. Through this procedure, a high-resolution magnetic field map can be predicted by measuring the fields at only $n$ sample locations.

	We validate the method using three simulated magnetic-field distributions with single, two, and three extrema, shown in Fig.~\ref{Fig2}(a,e,i). Measurements at each sampling point employ the sensitivity-enhanced PM-pulse protocol described above. Sampling locations are generated by perturbing uniformly spaced grid points with uniformly distributed displacements bounded to $30\%$ of the field-of-view size. With $n=25$, direct point-wise interpolation captures only coarse features (Fig.~\ref{Fig2}(b,f,j)), whereas Bayesian estimation reconstructs a $100\times100$ pixel map in good agreement with the ground truth (Fig.~\ref{Fig2}(c,g,k)).
	
\begin{figure}
		\centering
		\includegraphics[width=0.9\textwidth]{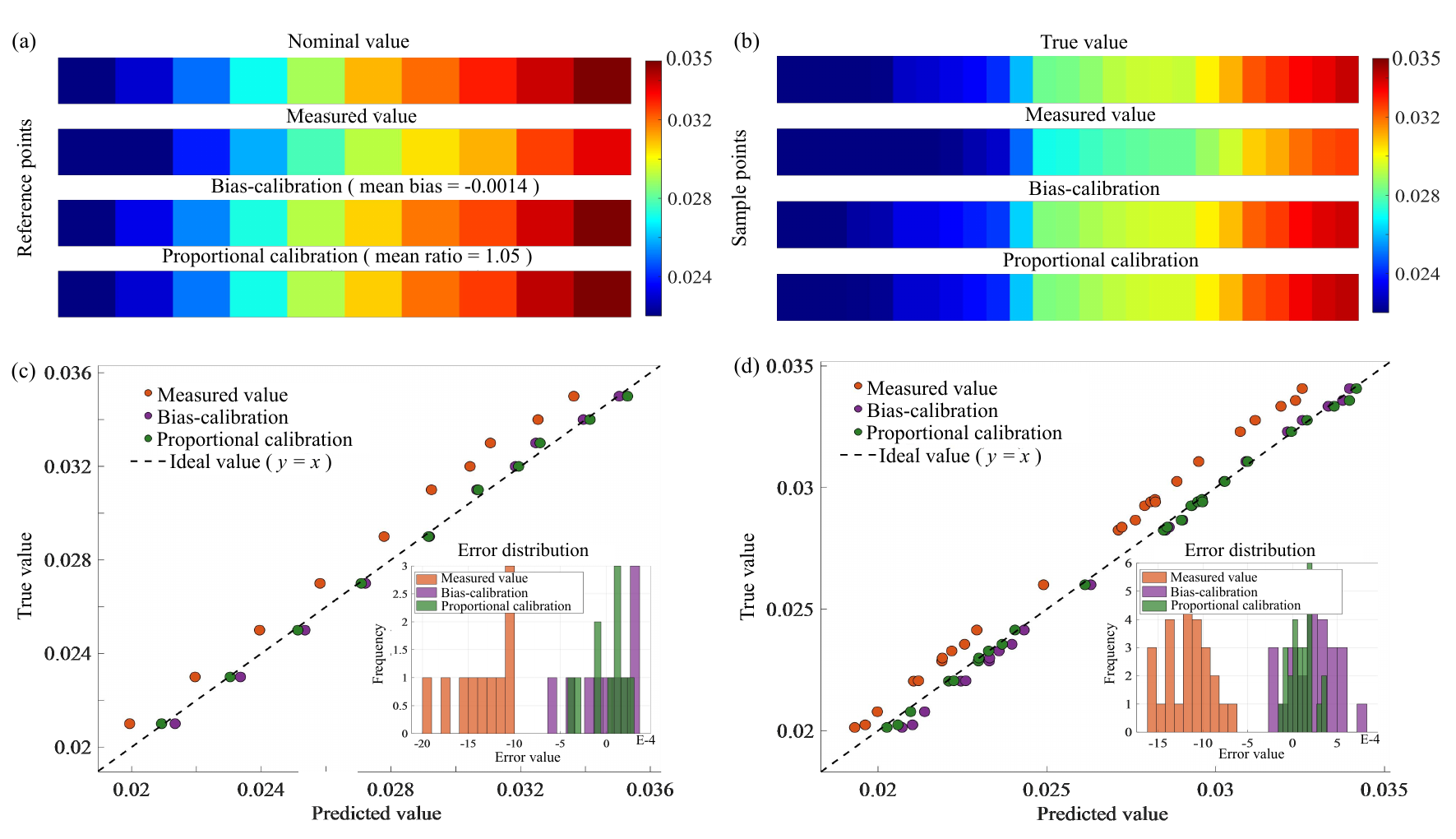}
		\caption{Illustration of the reference-based mean-adjustment procedure.
			(a) Values at the reference points, shown as nominal values, measured values, bias-calibrated values, and proportionally calibrated values. The reference values are uniformly chosen so as to span the range covered by the 25 sampling-point measurements.
			(b) Ground-truth values, unadjusted measurements, bias-calibrated values, and proportionally calibrated values for the 25 sampling points, where the calibration coefficients are obtained from the reference points.
			(c) Error comparison for the reference points before calibration, after bias calibration, and after proportional calibration.
			(d) Error comparison for the sampling points before calibration, after bias calibration, and after proportional calibration. For both the reference and sampling points, proportional calibration produces error distributions that are more tightly centered around zero than bias calibration.}
		\label{Fig3}
	\end{figure}

	Due to measurement errors at individual sampling points, the reconstructed global distribution can exhibit a residual systematic deviation from the ground truth. To mitigate this bias and improve the overall accuracy, we introduce a reference-based mean-adjustment procedure, which yields the improved reconstructions shown in Fig.~\ref{Fig2}(d,h,l). Specifically, ten reference points $\mathbf{r}$ with nominal values $y_{\mathrm{nom}}(\mathbf{r})$ were selected to be uniformly distributed so as to span the value range covered by the $25$ sampling-point measurements. These reference points were measured using the same protocol under identical noise conditions. The top two rows of Fig.~\ref{Fig3}(a) show the nominal and measured values at the reference points. The deviations between the measured and nominal reference values are then used to calibrate the measured data employed for image reconstruction.
	
	We denote the mean of the nominal and measured reference values by $\overline{y_{\mathrm{nom}}(\mathbf{r})}$ and $\overline{y(\mathbf{r})}$, respectively, and define the adjusted value at coordinate $\mathbf{x}$ as $y_{\mathrm{ad}}(\mathbf{x})$. We consider two calibration schemes. The bias-calibration approach applies an additive correction,
	\begin{equation}
		y_{\mathrm{ad}}(\mathbf{x}) = y(\mathbf{x}) + \left(\overline{y_{\mathrm{nom}}(\mathbf{r})}-\overline{y(\mathbf{r})}\right),
	\end{equation}
	whereas the proportional calibration rescales the data multiplicatively,
	\begin{equation}\label{Eq_prop-cali}
		y_{\mathrm{ad}}(\mathbf{x}) = y(\mathbf{x})\frac{\overline{y_{\mathrm{nom}}(\mathbf{r})}}{\overline{y(\mathbf{r})}}.
	\end{equation}
	The resulting calibrated reference values for both schemes are shown in the bottom two rows of Fig.~\ref{Fig3}(a). Compared with bias calibration, proportional calibration yields smaller absolute errors and a narrower error spread for the reference points, as shown in Fig.~\ref{Fig3}(c). Fig.~\ref{Fig3}(b) summarizes, for the $25$ sampling points, the ground-truth values, the unadjusted measurements, and the values obtained after bias and proportional calibration; the corresponding errors are shown in Fig.~\ref{Fig3}(d). These results verify the effectiveness of proportional calibration. Consequently, we adopt proportional calibration according to Equation (\ref{Eq_prop-cali}) to generate the adjusted values used for the MABE reconstructions in Fig.~\ref{Fig2}(d,h,l).
	
	\begin{figure}
		\centering
		\includegraphics[width=0.9\textwidth]{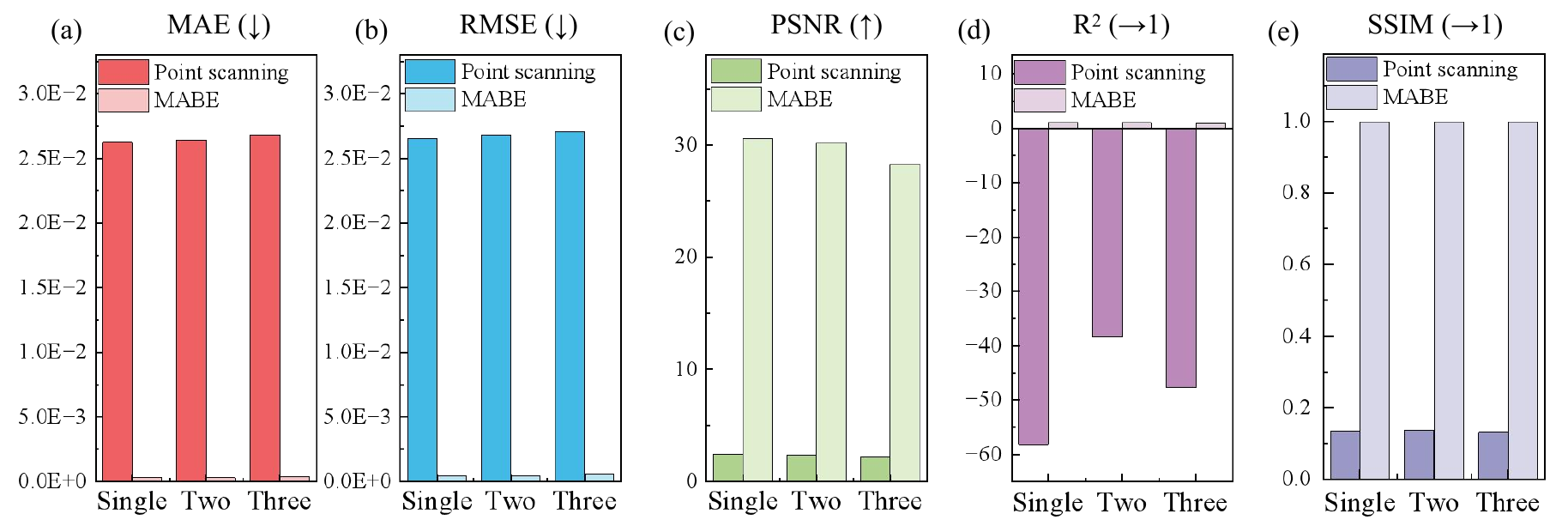}
		\caption{Quantitative image-quality metrics for the MABE reconstruction under irregular magnetic-field distributions containing single, two, and three extrema, as the sampling number $n$ varies:
			(a) mean absolute error (MAE);
			(b) root mean square error (RMSE);
			(c) peak signal-to-noise ratio (PSNR);
			(d) coefficient of determination ($R^2$);
			(e) structural similarity index (SSIM).
			In the figure captions, the arrows shown after each metric indicate whether smaller values ($\downarrow$), larger values ($\uparrow$), or values closer to unity ($\rightarrow 1$) correspond to better reconstruction performance.}
		\label{Fig4}
	\end{figure}
	
	Fig.~\ref{Fig4} quantitatively demonstrates the performance gain of the MABE method for the three magnetic-field distributions shown in the first column of Fig.~\ref{Fig2}, using five evaluation metrics~\cite{foxAppliedRegressionAnalysis2015,zhouwangImageQualityAssessment2004}: mean absolute error (MAE), which measures the average absolute deviation; peak signal-to-noise ratio (PSNR), which characterizes reconstruction fidelity relative to noise; root mean square error (RMSE), which weights larger errors more strongly; coefficient of determination ($R^2$), which summarizes goodness of fit; and structural similarity index measure (SSIM), which evaluates structural similarity. In general, smaller MAE and RMSE, together with larger PSNR and values of $R^2$ and SSIM closer to unity, indicate better reconstruction quality. Negative values of $R^2$ may occur when the reconstruction fits the data worse than the trivial predictor given by a horizontal line at the mean of the observed data~\cite{gonzalezDigitalImageProcessing2017,Kvalseth1985Cautionarynote2}.
	
	For magnetic-field distributions containing single, two, and three extrema, the proposed MABE approach yields substantial improvements compared with images obtained directly from $25$ sampling-point measurements. Specifically, MAE and RMSE are reduced from the $10^{-2}$ level to the $10^{-4}$ level. PSNR increases from approximately $2.3~\mathrm{dB}$ to $30.58~\mathrm{dB}$, $30.21~\mathrm{dB}$, and $28.3~\mathrm{dB}$, respectively. Meanwhile, $R^2$ increases from negative values to $0.983$, $0.988$, and $0.978$, and SSIM improves from about $0.13$ to $0.9998$. For all three magnetic-field distributions, the five metrics consistently indicate high-fidelity image reconstruction.
	
	\begin{figure}
		\centering
		\includegraphics[width=0.9\textwidth]{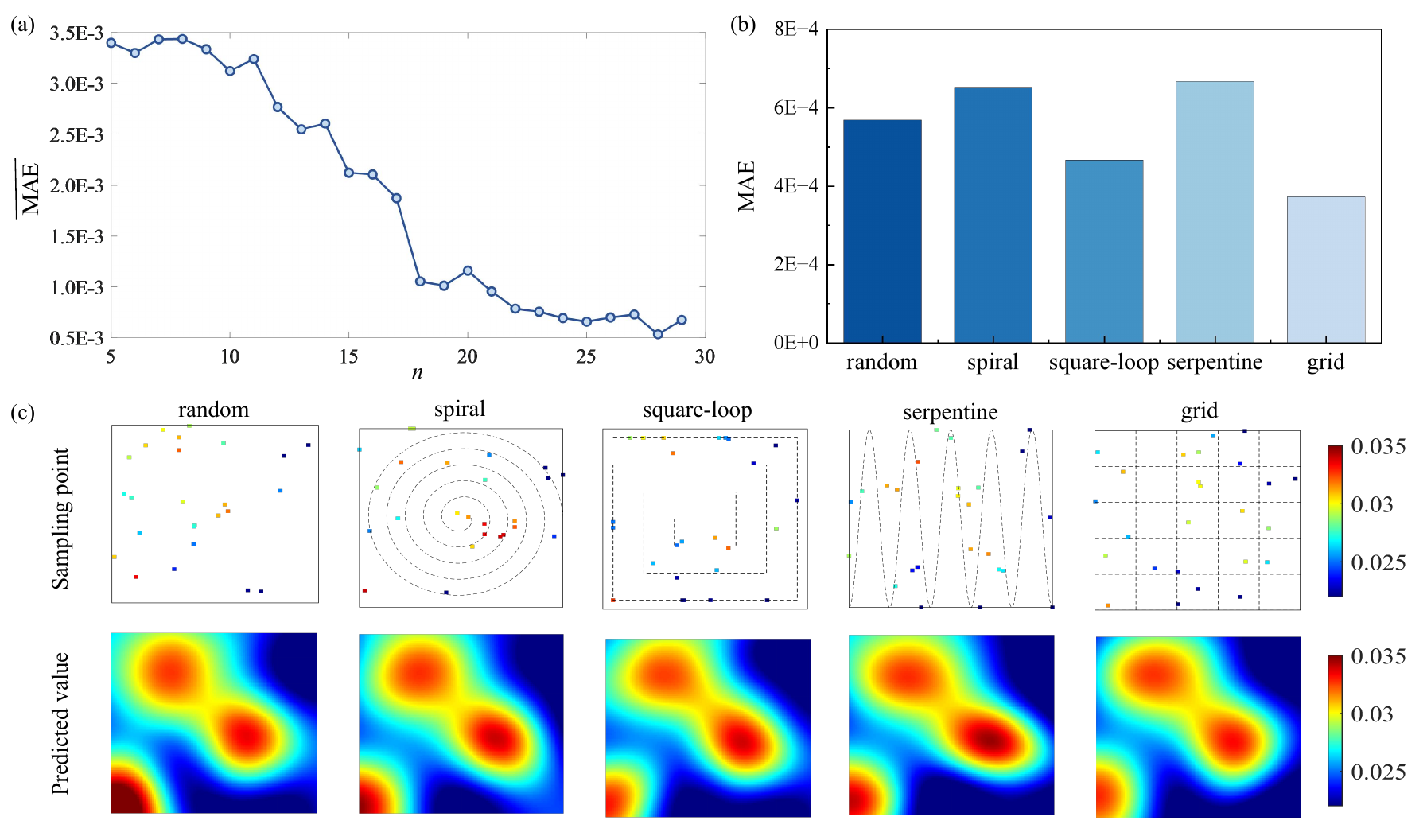}
		\caption{(a) Mean absolute error (MAE) over all $10^{4}$ pixels of the reconstructed image as a function of the number of sampling points used in the MABE method. Each data point represents the average over 20 independent reconstructions.
			(b) MAE of reconstructed images obtained using MABE under five sampling strategies (see the main text).
			(c) Sampling-point locations for the five strategies and the corresponding reconstructed $10^{4}$-pixel images obtained using MABE.}
		\label{Fig5}
	\end{figure}
	
	\begin{figure}
		\centering
		\includegraphics[width=0.9\textwidth]{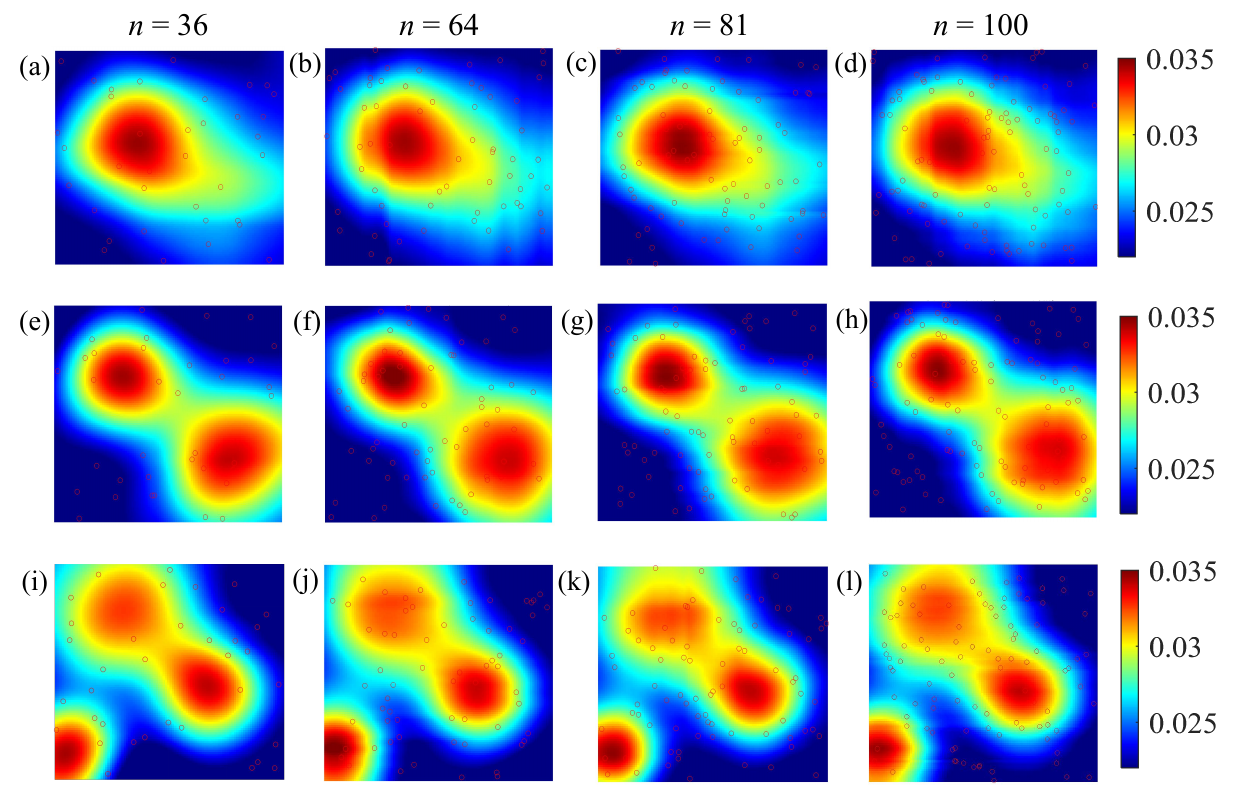}
		\caption{Reconstructed magnetic-field maps for distributions containing single, two, and three extrema using sampling numbers $n=25,36,64,81,$ and $100$, respectively. The corresponding ground-truth magnetic-field distributions are shown in Fig.~\ref{Fig2}(a,e,i).}
		\label{Fig6}
	\end{figure}
	
	\begin{figure}
		\centering
		\includegraphics[width=0.9\textwidth]{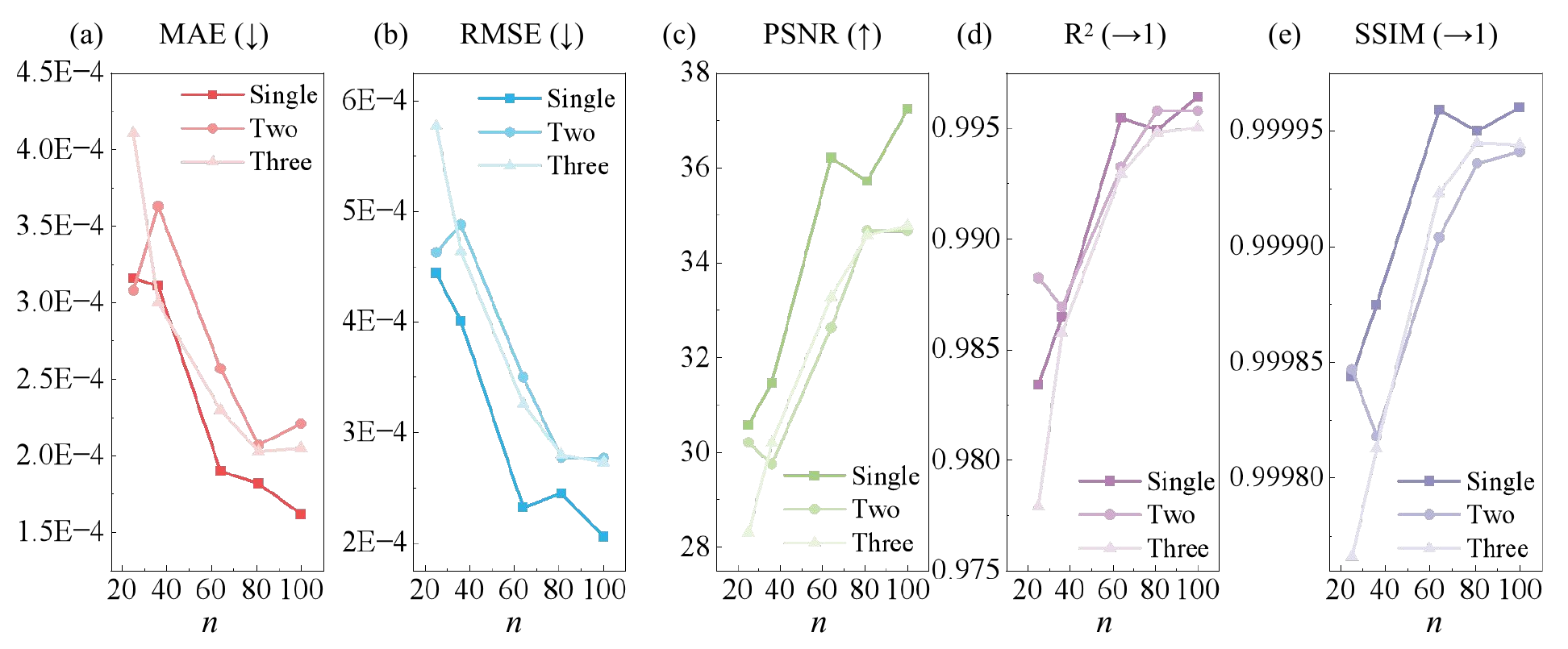}
		\caption{Quantitative image-quality metrics of the MABE reconstruction as the number of sampling points increases from 25 to 100:
			(a) mean absolute error (MAE);
			(b) root mean square error (RMSE);
			(c) coefficient of determination ($R^2$);
			(d) peak signal-to-noise ratio (PSNR);
			(e) structural similarity index (SSIM).
			Upward arrows in the titles indicate metrics for which larger values correspond to better performance, whereas downward arrows indicate metrics for which smaller values are better.}
		\label{Fig7}
	\end{figure}
	
	Fig.~\ref{Fig5} illustrates the influence of sampling number and sampling strategy, using the magnetic-field distribution with three extrema as an example. Fig.~\ref{Fig5}(a) shows the average MAE as a function of the number of sampling points, where each data point represents the mean over 20 independent MABE reconstructions with different sampling sets. For square numbers of sampling points, locations are generated by adding normally distributed random perturbations to uniformly spaced grid points, with the displacement magnitude bounded to $30\%$ of the field-of-view size; any remaining sampling points are drawn uniformly at random over the full field of view. As shown in Fig.~\ref{Fig5}(a), 25 sampling points are sufficient to maintain MAE stably at the $10^{-4}$ level. Balancing acquisition time and reconstruction fidelity, we therefore identify 25 sampling points as a practical compromise.
	
	Fig.~\ref{Fig5}(b) compares MAE obtained using five sampling strategies. The first strategy (``random'') draws sampling points uniformly at random across the field of view. The next three strategies generate points uniformly along spiral, square-loop, and serpentine trajectories, respectively, with additional uniformly distributed perturbations bounded to $30\%$ of the field-of-view size. The final strategy (``grid'') applies bounded uniform perturbations to the center coordinates of a uniform grid, as described above. Fig. ~\ref{Fig5}(c) provides a visual comparison of sampling locations and the corresponding reconstructed magnetic-field images. Among the five strategies, the grid-based strategy yields the best reconstruction performance and is therefore adopted as the default sampling scheme for the MABE method.
	
	Fig.~\ref{Fig6} and~\ref{Fig7} show the reconstructed magnetic-field images and the corresponding quantitative metrics obtained using MABE as the sampling number increases from 25 to 36, 64, 81, and 100. With increasing sampling density, the reconstructed boundary features become progressively more consistent with the ground truth, and all evaluation metrics exhibit an overall improving trend. When 100 sampling points are used, SSIM exceeds $0.9999$ for all three distributions. Notably, although the number of sampling points increases by a factor of four from 25 to 100, the corresponding improvement in the quantitative metrics is relatively modest. In practical applications, the trade-off between sampling speed and reconstruction quality should therefore be carefully considered.

	\section{Discussion}
	
	In this work, we have introduced a training-free reconstruction framework enabling widefield NV magnetometry from a highly sparse set of sampling points. In representative simulations, magnetic-field images containing $10^{4}$ pixels were successfully reconstructed from as few as $25$ measurement points, achieving structural similarity index values exceeding $0.999$. We further examined how sampling strategy and sampling density influence reconstruction performance. Compared with supervised learning--based approaches, the proposed method is training-free, structurally simple, and capable of maintaining good global prediction performance.
	
	It should be emphasized that the high reconstruction fidelity observed at low sampling densities is primarily achieved for relatively smooth and structurally simple magnetic-field distributions. For more complex spatial patterns, reliable reconstruction is expected to require higher sampling densities or the incorporation of more sophisticated inference strategies. In particular, approaches inspired by compressed sensing and Gaussian-process-aided adaptive sampling~\cite{kelleyFastScanningProbe2020,checaHighspeedMappingSurface2023,yangSpacefillingScanPaths2018,liAutomaticSparseESM2020} may provide promising routes to incorporate stronger prior information and further enhance reconstruction quality.
	
	In parallel, deep-learning-based reconstruction methods have demonstrated strong capability in restoring complex structures in medical imaging and super-resolution recovery of physical fields~\cite{sunDegradationAwareDeepLearning2021,sunEfficientDualdomainDeep2024,wuDualDomainDeepPrior2025,braureConditioningGenerativeLatent2025,collinsSuperresolutionUncertaintyEstimation2023}. While many of these methods rely on large supervised training datasets, others---such as deep image prior approaches~\cite{wuDualDomainDeepPrior2025}---exploit implicit network priors without explicit training data. Although their applicability to quantitative NV magnetometry remains to be fully assessed, such methods offer intriguing possibilities for accelerating widefield imaging and addressing complex current distributions in integrated circuits and related systems.

	\section*{Acknowledgments}
	The authors thank Yaoxing Bian, Shuangping Han, Jiamin Li, Kai Song and Hongrui Liu for their valueble discussions and supports.

	\section*{References}
	%参考文献
	%\clearpage %强制换页
	\renewcommand{\refname}{} % 清空article类内置标题
	\bibliographystyle{unsrt}
	\bibliography{ref}

\end{document}